\documentclass[twocolumn,showpacs,preprintnumbers,amsmath,amssymb,superscriptaddress]{revtex4-1}
\usepackage{graphicx}% Include figure files
\usepackage{dcolumn}% Align table columns on decimal point
\usepackage{bm}% math
\usepackage{color}% color
\usepackage{pifont}
\usepackage{natbib}
\usepackage{hyperref}
\hypersetup{
colorlinks = true,
urlcolor   = blue,
linkcolor  = blue,
citecolor  = blue
}

\begin{document}

\preprint{APS/PRB}
\title{Li-ion diffusion in single crystal LiFePO$_4$ measured by muon spin spectroscopy}

\author{Ola~Kenji~Forslund}
 \email{okfo@kth.se}
\affiliation{Department of Applied Physics, KTH Royal Institute of Technology, SE-106 91 Stockholm, Sweden}
\author{Rasmus~Toft-Petersen}
\affiliation{Department of Physics, Technical University of Denmark, DK-2800, Kgs. Lyngby, Denmark}
\author{David~Vaknin}
\affiliation{Ames Laboratory and Department of Physics and Astronomy, Iowa State University, Ames, Iowa 50011, United States}
\author{Natalija van Well}
\thanks{\emph{Department of Earth and Environmental Sciences, Crystallography Section, Ludwig-Maximilians-University Munich, D-80333, Munich, Germany}}
\affiliation{Laboratory for Neutron Scattering \& Imaging, Paul Scherrer Institute, CH-5232, Villigen, PSI, Switzerland}
\author{Mark~Telling}
\affiliation{ISIS Facility, Rutherford Appleton Laboratory, Chilton, Didcot Oxon OX11 0QX, United Kingdom}
\author{Yasmine Sassa}
\affiliation{Department of Physics, Chalmers University of Technology, SE-412 96 G\"oteborg, Sweden}
\author{Jun~Sugiyama}
\affiliation{CROSS Neutron Science and Technology Center, Tokai, Ibaraki 319-1106, Japan}
\author{Martin~M\aa nsson}
\affiliation{Department of Applied Physics, KTH Royal Institute of Technology, SE-106 91 Stockholm, Sweden}
\author{Fanni~Juranyi}
\affiliation{Laboratory for Neutron Scattering \& Imaging, Paul Scherrer Institute, CH-5232, Villigen, PSI, Switzerland}

\date{\today}

\begin{abstract}
Muon spin spectroscopy ($\mu^+$SR) is now an established method to measure atomic scale diffusion coefficients of ions in oxides. This is achieved via the ion hopping rate, which causes periodic change in the local magnetic field at the muon site(s). We present here the first systematic study on a single crystalline sample. The highly anisotropic diffusion of Li-ions in the battery cathode material LiFePO$_4$, combined with the extensive investigation of this material with $\mu^+$SR and other techniques make it a perfect model compound for this study. With this experiment we can confirm that Li diffusion in the bulk LiFePO$_4$ is measurable with $\mu^+$SR. Hence, surface/interface effects, which might play a crucial role in case of powders/nano crystals, are less significant for macroscopic single crystals where bulk diffusion is in fact present. We observe that the internal magnetic field fluctuations caused by the diffusing Li-ions are different depending on the crystal orientation. This is not obviously expected based on theoretical considerations. Such fluctuation rates were used to estimate the diffusion coefficient, which agrees well with values estimated by first principle calculations considering anisotropic diffusion. 

%In particular, we have shown that powder measurements on compounds with anisotropic ion diffusion inevitably reduces the estimated diffusion coefficient.

\end{abstract}



\maketitle

\section{\label{sec:Intro}Introduction}
The diffusion of Na and Li ions in battery cathode materials is governed by the fundamental parameter, the diffusion coefficient ($D_{\rm Li,Na}$). Such parameter is conventionally determined by electrochemical measurements and have contributed significantly to the battery boom seen in the 21$^{\rm st}$~century. The derived $D_{\rm Li,Na}$ usually have a spread of several orders in magnitude, due to the fact that the reactive surface area of a porous liquid electrode is nearly impossible to estimate accurately. Moreover, different methods probe diffusion over different distances, where beside bulk diffusivity other effects, like interfaces can also play a role, and sometimes it is impossible to separate the different contributions. \cite{Yao1995, Sugiyama2013, Dokko2001} Instead, microscopic probes are needed and Li/Na-NMR have successfully manage to fill this gap \cite{Grey2004, Heitjans2005}. However, cathode materials with transition metal contains $d$-electron spins that contributes to the spin-lattice relaxation rate (1/T$_1$). Therefore, accurately determining absolute values of the ion diffusion coefficient is quite often challenging \cite{Nakamura2000, Van2000, Graham2016}. 

Recently, another microscopic probe capable of determining the diffusion coefficient on the atomic scales has appeared \cite{Sugiyama2009}. Muon spin rotation, relaxation and resonance ($\mu^+$SR) relies on implantation of muons to interact with the local magnetic field in the compounds. The muon, with its high gyromagnetic ratio, can detect even nuclear moments and fluctuations of such. Therefore, the muon is a small magnetic probe capable of detecting magnetic field fluctuations caused by a diffusing spices with a nuclear moment, such as Li or Na with $I_{Li,Na}=3/2$. Indeed, early measurements on LiCoO$_2$ suggested a diffusion coefficient comparable to those predicted by first principle calculations \cite{Sugiyama2009}. Since then, many Na and Li cathode materials has been measured by means of $\mu^+$SR and intrinsic values of $D_{Li,Na}$ have been determined \cite{Sugiyama2010, Sugiyama2013_2, Mansson2013}. Notably, the conversion of magnetic field fluctuation into $D$ requires the information of the diffusion paths of the diffusing spices. Experimentally, neutron diffraction can be used to determine these paths. In case of LiFePO$_4$, maximum entropy analysis combined with neutron diffraction experimentally verified a quasi-1D diffusion of Li ions \cite{Nishimura2008}. The quasi-1D behaviour was also confirmed by AC impedance spectroscopy measurements on single crystals \cite{Jiying2008}.

\begin{figure*}[ht]
  \begin{center}
    \includegraphics[keepaspectratio=true,width=0.95\textwidth]{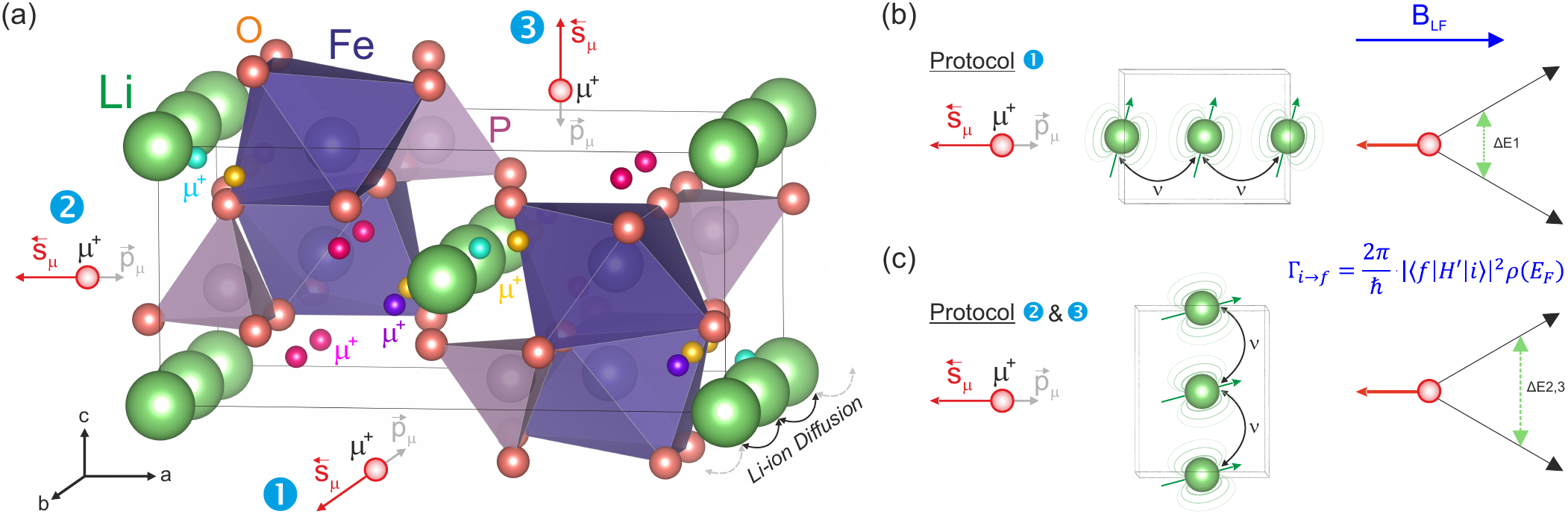}
  \end{center}
  \caption{(a) Atomic structure of LiFePO$_4$ the listed possible muon sites (table~\ref{table:table1}) and we define three measurements protocols; protocol~1 when $S_{\mu}\parallel b$, protocol~2 when $S_{\mu}\parallel a$ and protocol~3 when $S_{\mu}\parallel c$. Protocol~1 is highlighted in (b) while protocol~2 and 3 are highlighted in (c). The energy diagram in (b) and (c) schematically shows the Zeeman split of the muon state due to the applied longitudinal field ($B_{\rm LF}$), for which the local field fluctuation and thus the energy supply is smaller in (b) ($\Delta E_{1}$) than in (c) ($\Delta E_{1,2}$), given the geometry, see main text. The transition probability per unit of time from initial ($i$) and final states ($f$) is given by Fermi's golden rule ($\Gamma_{i,f}=\frac{2\pi}{\hbar}\mid<f\mid H'\mid i>\mid ^2 \rho(E_{\rm f})$).
  }
  \label{fig:Structure}
\end{figure*}

In this letter, we report first $\mu^+$SR experiment of ion diffusion in a single crystal. The muon time spectrum associated to Li diffusion in an isotropic ensemble (e.g. powder samples) can be described by the  dynamical Kubo-Toyabe function \cite{Sugiyama2009}. In general, only transverse internal magnetic field components contribute to the resulting field distributions at the muon site. In detail, fluctuating transverse field components are considered perturbative whereas the longitudinal field components only affect the degree of the Zeeman split. Since the nuclear magnetic fields are dipoles, the number of nuclear magnetic field components contributing to the transverse and longitudinal directions, with respect to the muon spin, depends on the relative position of the Li ion. Similarly, the degree on which the local magnetic field at the muon site changes because of diffusing Li-ions depends thus on the specific diffusion directions. Therefore, different values of the spin-lattice relaxation rate can be obtain depending how the initial muon spin polarisation is oriented with respect to the expected Li-ion diffusion path. Naturally, such difference is most visible in low dimensions and single crystal LiFePO$_4$ was chosen in our first study with its distinct quasi-1D Li diffusion path. 

%In other words, if the initial muon spin polarisation is parallel to the Li-ion diffusion path, the perturbation field is smaller and so the probability of a moun spin flip is reduced. 

\section{\label{sec:exp}Experimental Setup}
Single crystals of LiFePO$_4$ were grown at Ames Lab using the flux method as described in \cite{Li2006}. The crystal structure (and magnetic properties) are extensively described in \cite{Rasmus2012}. Two pieces of single crystals (4 x 5 x 6 mm$^3$ and 2 x 3 x 6 mm$^3$) were aligned and mounted on a newly designed silver sample holder for the $\mu^+$SR measurements, which were performed at the EMU beamline at ISIS. Due to the small size of the crystals it was crucial to minimize the amount of silver in the beam and to use the flypast mode. Since the orientation of the crystals had to be changed, both crystal pieces were glued on a small, rectangular silver piece such that the crystal axis were parallel to the sides of the silver piece. These were then fixed on the silver sample holder, and were covered with Al foil in order to minimise the Temperature gradient in the sample. The temperature was controlled by a CCR cryostat. The crystals were investigated in active zero field (true zero field condition by compensating for Earth's magnetic field) and in transverse and longitudinal fields. Here, transverse and longitudinal field refers to the applied field direction with respect to the initial muon spin direction. The samples were measured in three configurations, such that a, b and c-axis of the crystals were parallel to the initial muon spin polarisation, and are highlighted in the text as bc, ac and ab. Finally, the $\mu+$SR data was analysed using \texttt{musrfit} \cite{musrfit}. 

\begin{figure*}[ht]
  \begin{center}
    \includegraphics[keepaspectratio=true,width=0.85\textwidth]{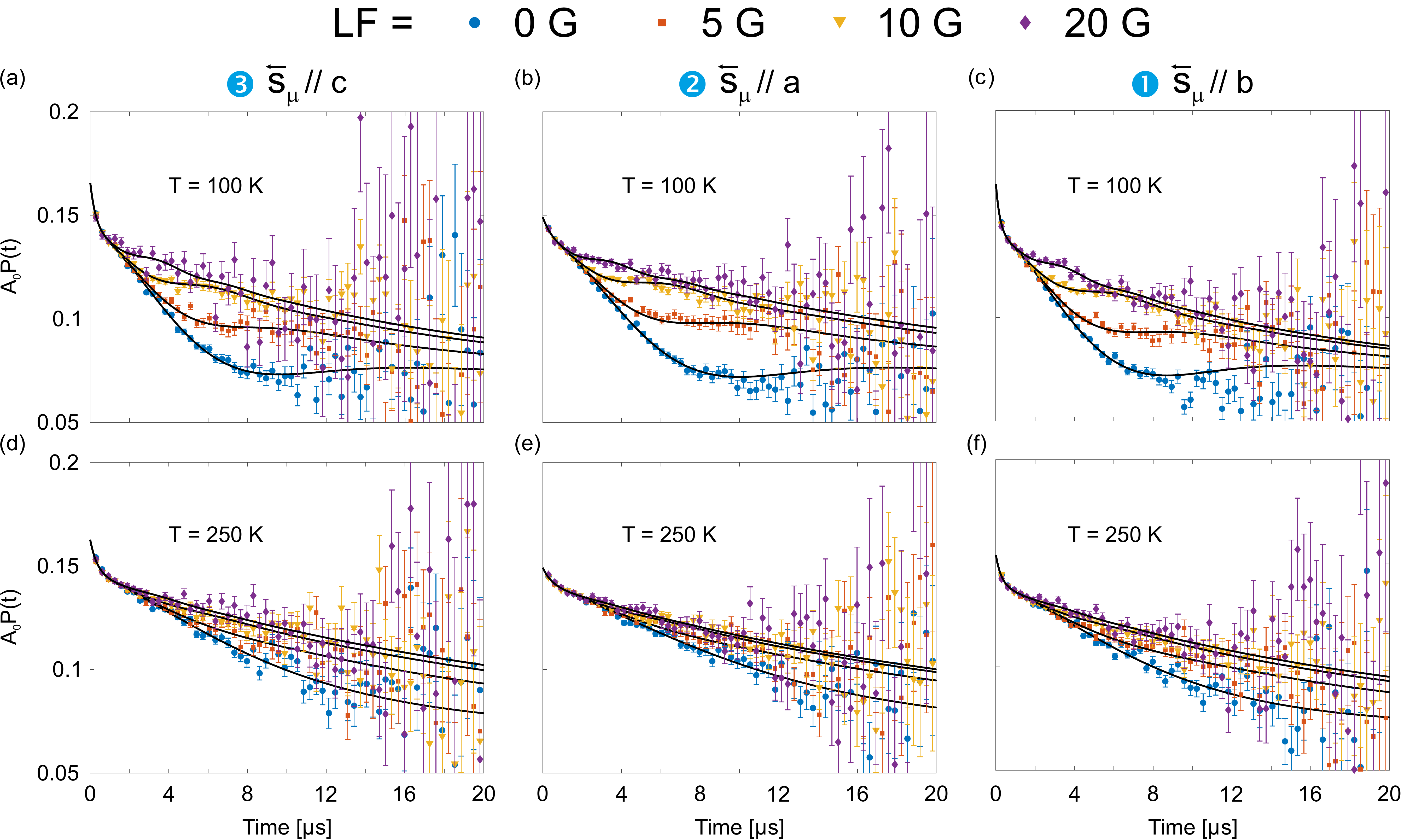}
  \end{center}
  \caption{The zero field (ZF) and longitudinal field (LF) $\mu^+$SR time spectra measured at $T=100~K$ (a,b,c) and $T=250$~K (d,e,f) for each measurement protocols; protocol~1 when $S_{\mu}\parallel b$ (c, f), protocol~2 when $S_{\mu}\parallel a$ (b, e) and protocol~3 when $S_{\mu}\parallel c$ (a, d) [see also Fig.~\ref{fig:Structure}]. The solid line corresponds to best fit obtained using Eq.~$\ref{eq:ZFLF}$.
  }
  \label{fig:_ZFLFSpec}
\end{figure*}

\section{\label{sec:theory}Theoretical background}
Li (jump) diffusion in $\mu^+$SR experiments can be treated within the so called strong collision model. It assumes a static field (at the muon site) until the local field is changed suddenly due to the Li dynamics, and the local field takes on a new value according to a Gaussian-Markovian process. Such process is solved analytically in Laplace space \cite{Yaouanc2011}

\begin{equation}
F(s)=L(P_{dyn})=\frac{f_0(s)}{1-\nu f_0(s)}
\label{eq:DynamicKT}
\end{equation}

where $F(s)$ is the Laplace transform of $P_{\rm dyn}$, which is the sought dynamical polarisation function. $f_0(s)$ on the other hand is the Laplace transform of the static polarisation function. The dynamical polarisation function depends solely on the static polarisation function and on the spin-lattice relaxation rate, $\nu$, $i.e.$ on the hopping rate.Therefore, the static polarisation function and $\nu$ need to be evaluated. Detailed derivation of Eq.~\ref{eq:DynamicKT} can be found elsewhere \cite{Yaouanc2011}.

The dipolar coupling between the muon and the Li nuclear moment depends on the specific distance and coordinate of the muon site. The muon sites for this compound was evaluated in the past \cite{Sugiyama2011}, and the expected internal field distributions at these sites can be computed in the Van Vleck limit. The values of powder average and along the crystallographic directions for each muon site are listed in table~\ref{table:table1}. Given that the computed $\Delta$ values are quite similar for all directions, the static polarisation function is expected to be given by the Gaussian Kubo-Toyabe (G-KT) function even for the single crystal. Such assertion is supported by the quality of the fit (Fig.~\ref{fig:_ZFLFSpec}) and is also further discussed below. 

With the static polarisation function determined, the question that needs to be answered is how the value of $\nu$ is determined. The spin-lattice relaxation rate, the muon spin transition rate between the two Zeeman states, is given by the Fermi's Golden rule based on the perturbating field, $i.e.$ the fluctuating field. The pertubating Hamiltonian is $H_{\rm per}=-\frac{1}{2}\gamma_{\mu}\hbar\bm \sigma \cdot \delta \bm B$ and is thus given by the magnitude of the fluctuations. Moreover, these fluctuating fields need to be transverse to the muon spin in order to transfer the required energy, determined by the degree of the Zeeman-split which in turn depends on the strength of the parallel field component. However, we shall neglect the effect of parallel field component on the transition rate, and instead assume that the transition rate is given solely on the magnitude of the perpendicular field components. 

Based on the relative position of the Li and the muon, it is easy to see that number of dipole field components contributing to the perpendicular local field is lower in cases the Li is situated parallel to the muon spin, as opposed to when the Li is situated perpendicular with respect to the muon spin. Naturally, this affects the value of $\delta \bm B$ and thus the muon transition probability, $i.e.$ the hopping rate $\nu$. This assertion is also supported by the data presented below.

Finally, since the static polarisation function is the G-KT, the dynamical polarisation function, $P_{\rm dyn}$, is simply the dynamic GKT (DGKT). Additionally, the response of a muon ensemble originating from two independent magnetic field sources is given by the Fourier transform of their convolution. Therefore, the total repose can be modeled as a product of an a dynamic GKT (DGKT), originating from a fluctuating nuclear magnetic fields, and an exponential relaxation, originating from a fluctuating Fe moments. The overall fitting function is given by Eq.~\ref{eq:ZFLF} where the components are described in detail below. 

%similar Delta for different directions
%Explain/derive local magnetic field fluctuation. 
%Justify the fit function: hopping rate does not depend on the direction of diffusion; 
%Recall possible muon sites, 

\section{\label{sec:results}Results}
The diffusive behaviour of Li in LiFePO$_4$ was measured under ZF and LF = 5, 10 and $20~$Oe. Figure~$\ref{fig:_ZFLFSpec}$ show the measured $\mu^+$SR time spectrum collected at $T=100$ and $T=250$~K for the three different crystal orientation: protocol~1 ($S_{\mu}\parallel b$-axis), protocol~2 ($S_{\mu}\parallel a$-axis) and protocol~3 ($S_{\mu}\parallel c$-axis) where $S_{\mu}$ is the initial muon spin polarisation. At 100~K, a clear Gaussian Kubo-Toyabe (GKT) behavior is seen, like many paramagnetic compounds. As the temperature increases, the DGKT behaviour is evolved into a more exponential like behaviour, most clearly visible for protocol~2 but also for the protocol~3. On top of that, an initial fast relaxing signal is seen together with a small offset, a contribution of the sample holder. Therefore, the ZF+LF time spectra were fitted using a none relaxing background, a dynamic Gaussian Kubo-Toyabe (DGKT) and a fast relaxing component:

\begin{eqnarray}
 A_0 \, P_{\rm ZFLF}(t) &=&
A_{\rm KT}G^{DGKT}(H_{\rm LF},\Delta,\nu,t)e^{-\lambda}
\cr
 &+& A_{\rm F}e^{-\lambda_{\rm F} t}+A_{\rm BG},
\label{eq:ZFLF}
\end{eqnarray}

where $A_0$ is the initial asymmetry determined by the instrument and $P_{\rm ZFLF}$ is the muon spin polarisation under ZF and LF. $A_{\rm KT}$, $A_{\rm BG}$ and $A_{\rm F}$ are the asymmetry resulting from each magnetic environment the muon resides in: $A_{\rm KT}$ is the fraction of muons depolarising from nuclear moments, $A_{\rm BG}$ is the fraction of muons stopping outside the sample (mainly the sample holder) and $A_{\rm F}$ is the fraction of muons sensing Fe $d$-moment fluctuations. $\Delta$ is the supposed field distribution width of the internal Gaussian distribution, originating mainly from $I_{\rm Li}=3/2$, and $\nu$ is the fluctuation of of such internal nuclear field. $\lambda$ and $\lambda_{\rm F}$ are exponential relaxation rates, previously determined to be local Fe moment fluctuations \cite{Sugiyama2011}. 

\begin{figure}[ht]
  \begin{center}
    \includegraphics[keepaspectratio=true,width=85 mm]{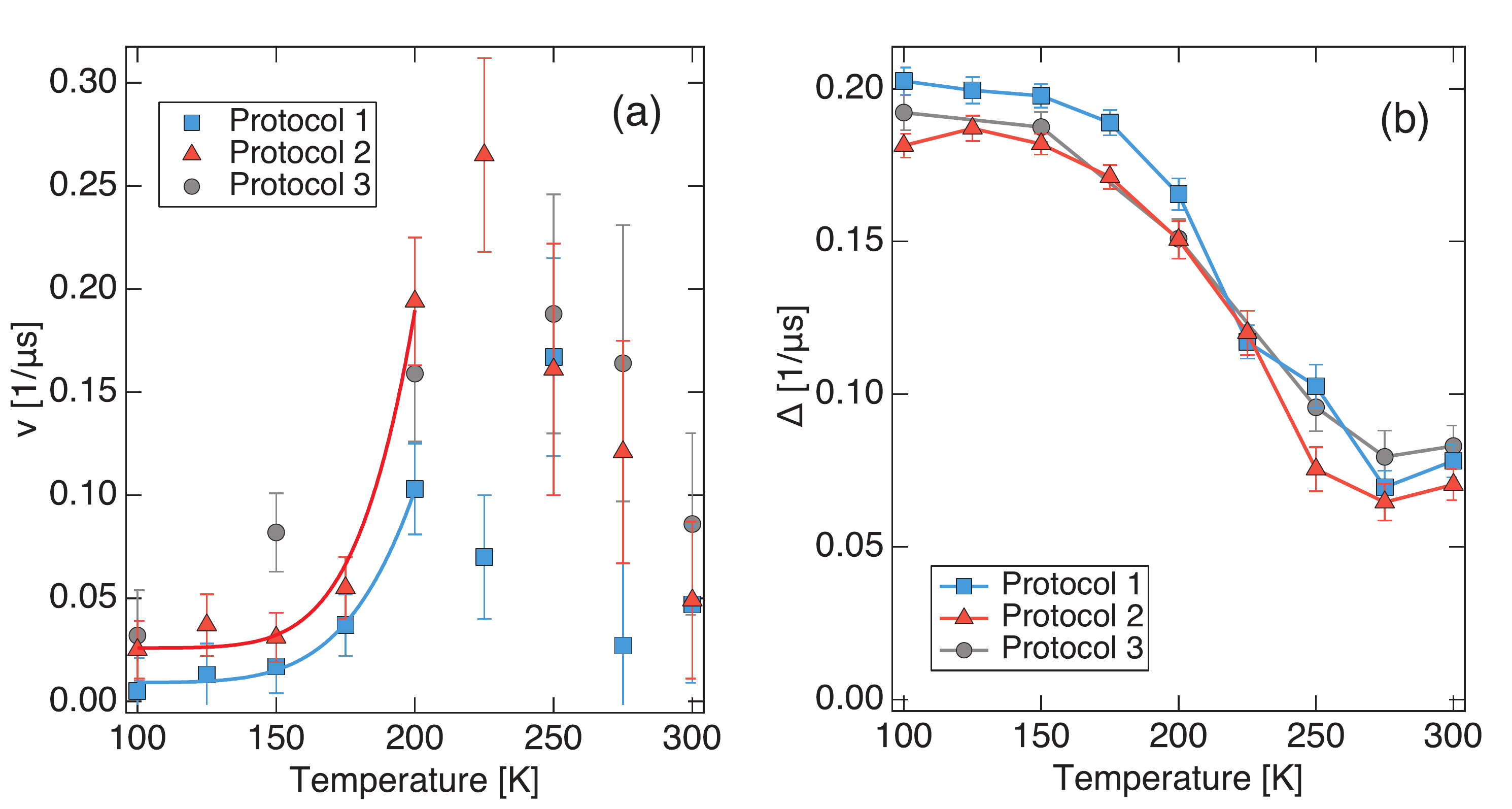}
  \end{center}
  \caption{Temperature dependence of (a) field fluctuation rate ($\nu$) and (b) field distribution width ($\Delta$) obtained for each measured protocol (protocol~1: $S_{\mu}\parallel b$, protocol~2: $S_{\mu}\parallel a$ and protocol~3: $S_{\mu}\parallel c$). The solid lines in (a) are best fits using $\nu=Ae^{-E_{\rm a}/k_{\rm B}T}+\nu_0$ up to $T=200$~K.
  }
  \label{fig:_Para}
\end{figure}

Figure~$\ref{fig:_Para}$ show temperature dependencies, from 100 to 300~K, of $\Delta$ and $\nu$ obtained using Eq.~(\ref{eq:ZFLF}). $A_{\rm BG}$ was estimated from a weak transverse field measurement conducted at $T<T_{\rm N}=50$~K for protocol~2 and 1 measurements while a $T=100$~K measurement was instead used for the ab direction. At 100~K, all three directions yield similar results, namely $\Delta\simeq0.2~\mu$s$^{-1}$ and $\nu\simeq0~\mu$s$^{-1}$. The value of $\Delta$ at 100~K is close to the obtained value from powder \onlinecite{Sugiyama2011}. As the temperature increases, $\Delta$ is steadily lowered from about 150~K until a new value is stabilised from about 250~K. These new values ($\sim0.07$) are significantly lower than the predicted ones in the static regime (table~\ref{table:table1}) and suggest dynamical origin behind the reduction. Such behaviour has been observed in many Li and Na cathode materials, and has been ascribed as motional narrowing \cite{Sugiyama2009, Sugiyama2010, Forslund2020_Na}. Since $\Delta$ corresponds roughly to the spin-spin relaxation rate, a similar temperature dependence is observed in the linewidth of LiCoO$_2$ measured with Li-NMR \cite{Nakamura2000} and also in weak transverse field relaxation rate (not shown). The saturation of $\Delta$ observed above $T\simeq250$~K is thus understood from the fact that $\mu^+$SR is only sensitive to specific time window. In fact, the high temperature value of $\Delta$ is consistent with the calculated field distribution width of Li$_0$FePO$_4$ and supports the scenario of motional narrowing to rapidly diffusing Li ions.

$\nu$ show $\nu\simeq0~\mu$s$^{-1}$ at 100~K, as expected for the static case. As temperature increases from about 150~K, just when $\Delta$ starts to decrease, an increase in $\nu$ is seen in all directions. That being said, the most prominent increase is seen in protocol~2 and 3, consistent with Li diffusion along the $b$-axis. As mentioned, these protocols represents a configuration in which the Li diffusion path (the $b$-axis) perpendicular to the initial muon spin. A small yet notable increase of $\nu$ is still observed in the protocol~1 (Li diffusion path parallel to the initial muon spin). As described above, perpendicular field components are always presents independent of the diffusion direction with respect to the muon spin, which inevitably yield muon spin-lattice relaxation. However, the magnitude of the pertubating field changes and so will the transition probability (and thus the size of $\nu$). Practically, the diffusion is quasi-1D, suggesting that the simple picture given above is actually smeared out. %Moreover, a small misalignment between the initial muon spin and the crystal axis can be expected, since the initial muon spin polarisation is about $\simeq10^\circ$ tilted with respect to the horizontal plane of the instrument. 
The large spread of $\nu$ above $\simeq 225$~K can be explained by the limited time window of $\mu^+$SR. Fitting the $\mu^+$SR time spectrum from $T\simeq 225$~K becomes difficult since the DGKT is expected to have an exponential like polarisation in the motional narrowing limit. In other words, it is difficult to separate the DGKT contribution from the multiplied exponential, which is included to account for the additional relaxation from $d$-orbital electrons. As also seen in Fig.~$\ref{fig:_ZFLFSpec}$(e), the applied field is almost overlapping with the other fields and the ZF spectrum. 

Given the above discussions, the Li atom is expected to be thermally activated at elevated temperatures. An exponential like evolution of $\nu(T)$  (Arrhenius behaviour, $\nu(T)=Ae^{-E_{a}/k_{\rm B}T}+\nu_0$) is expected and also observed in LiFePO4 powders \cite{Sugiyama2011}. Although our data is very limited both regarding number of data points and statistics, we can obtain a rough value for the activation energy $E_{\rm a}$ from the temperature range of $T=100$~K to 200~K (solid lines in Fig.~\ref{fig:_Para}(a)). $k_{\rm B}$ is Boltzmann constant, $\nu_0$ is a small background contribution and $A$ is a prefactor, which is assumed to be temperature independent. Similar activation energies are obtained for both directions, $E^{\rm protocol~2}_{\rm a}=0.17(4)$~eV and $E^{\rm protocol~1}_{\rm a}=0.14(2)$~eV, that is also consistent with powder measurements \cite{Sugiyama2011}. 

%The increased temperature range in the fits, up to 225~K and 250~K, result not only in small activation energies but also relatively large difference between the two directions, which is unreasonable. Therefore, the discuss above regarding the limited time window seems reasonable. Although, regardless of fitting range, a clear increase (decrease) of $\nu$ is seen in the cb (ca) direction.

\begin{figure}[ht]
  \begin{center}
    \includegraphics[keepaspectratio=true,width=65 mm]{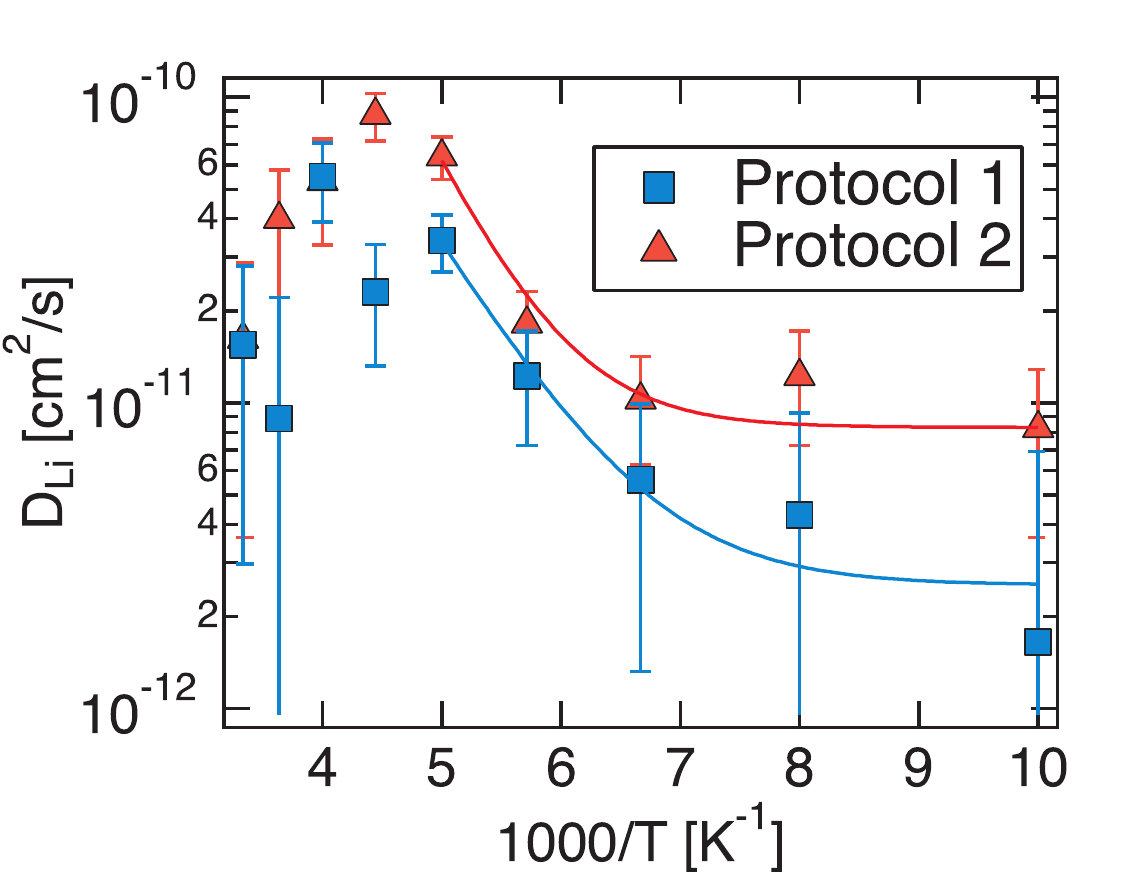}
  \end{center}
  \caption{Calculated diffusion coefficient ($D_{\rm Li}$) from values presented in Fig.~$\ref{fig:_Para}$, plotted in log scale vs inverse temperature for each measured protocol (protocol~1: $S_{\mu}\parallel b$, protocol~2: $S_{\mu}\parallel a$ and protocol~3: $S_{\mu}\parallel c$). The solid lines are fits from lowest to 200~K. The small upturn at lower temperatures are due to the small BG contribution $\nu_0$.
  }
  \label{fig:D}
\end{figure}

From the presented data, it is clear that the Li-diffusion is manifested predominantly along the $b$-axis. Since the Li diffusion path has been confirmed, we may now try to evaluate $D_{\rm Li}$ using presented data. Assuming that the field fluctuation rate $\nu$ can be considered as the Li hopping frequency between interstitial sites, the diffusion coefficient is given by

\begin{eqnarray}
D_{\rm Li}=\sum^{n}_{i=1}\frac{Z_{\nu,i}s^{2}_{i}\nu}{N_{i}}
\label{eq:D}
\end{eqnarray}

where $N_{i}$ is the number of Li sites in the $i$-th jump path, $Z_{\nu,i}$ is the vacancy fraction and $s_{i}$ is the jump distance. Following the calculation of Ref.~\onlinecite{Sugiyama2011} with $n=2$, $N_1=2$, $s_1=1.86$~\AA, $Z_1=1$ and $N_2=2$, $s_2=1.77$~\AA, $Z_2=1$, in which interstitial sites ((0.18, 0.98, 0) and (0.09, 0.75, 0)) are assumed to be the mediator of the hopping, diffusion coefficients as presented in Fig.~$\ref{fig:D}$ are obtained. Even though the $\mu^+$SR time window limits the measurement up to about 225~K, the diffusion coefficient can be estimated for 300~K by extrapolation the Arrhenius fits: $D^{\rm protocol~2}_{Li}(300~$K$)=1.23 \times 10^{-9}$~cm$^{2}$/s. Similar estimation from powder data yielded slightly lower values $D^{\rm powder}_{Li}(300~$K$)=3.6 \times 10^{-10}$~cm$^2$/s \cite{Sugiyama2011}. Of course, $D_{\rm Li}$ at 300~K were extrapolated for both the powder and single crystal measurements, and the obtained value depends on the quality of Arrhenius fit. %We thus note that $D^{\rm powder}_{Li}(200~$K$)\sim7 \times 10^{-11}$~cm$^2$/s \cite{Sugiyama2011} and $D^{\rm protocol~2}_{Li}(300~$K$)=6.4 \times 10^{-11}$~cm$^{2}$/s, which are indeed comparable. 
By considering the electrostatic structure, first principle calculation \cite{Morgan2004} proposed a value of $D^{\rm calc}_{\rm Li}=10^{-8}$~cm$^{2}$/s at 300~K. Such value is close to the extrapolated values based on our single crystal data, making them closest agreement with calculations yet. The agreement can be attributed to higher sample quality of single crystals (in comparison to powder). More importantly however, the fact that powder measures the average in all three directions will inevitably reduce the estimated value of the diffusion coefficient, at least in cases where the diffusion path is highly anisotropic.

\begin{table*}[ht]
\caption{\label{table:table1}
The calculated $\Delta$ values for powder and each crystalline direction for each proposed muon sites by Ref.~\cite{Sugiyama2011}. Please note, the last listed site is governed by the interaction with Fe, resulting into $A_{\rm F}$ of the measured signal \cite{Sugiyama2011}. Consequently, only the first three sites are sensing Li-ion hopping.
}

\begin{ruledtabular}
\renewcommand*{\arraystretch}{1.4}
\begin{tabular}{cccccc}
 muon site & powder [1/$\mu$s] & $\Delta^{\rm XX}$ [1/$\mu$s]& $\Delta^{\rm YY}$ [1/$\mu$s]& $\Delta^{\rm ZZ}$ [1/$\mu$s] &\\
 \hline
 (0.8146, 0.0404, 0.8914)& 0.204323 &0.219613 & 0.306436&0.306426&\\
  (0.3901, 0.2500, 0.3599)& 0.267609 &0.373966& 0.348544&0.405012&\\
   (0.0416, 0.2500, 0.9172)& 0.494660 &0.750251 & 0.538146&0.766887&\\
    (0.1225, 0.3772, 0.8679)& 0.365234&0.479855&0.558771&0.556163&\\
    
\end{tabular}
\end{ruledtabular}
\end{table*}

\section{\label{sec:discussion} Discussion}
The temperature dependencies of the spin-lattice relaxation rate, i.e. $\nu$, changes with the crystal orientation. As described above, the spin-lattice relaxation rate is given by the Fermi's golden rule based on the pertubative field, here the fluctuating field. Since the relative positions of the Li and the muon change the number of perpendicular field components, and thus the amplitude of the fluctuating field, the difference in the observed $\nu$ is ascribed not to the different jump probability of the Li ion in the different directions, but to the change in transition probability between the two Zeeman states. Therefore, the anisotropy of Li diffusion in bulk LiFePO$_4$ has been approved experimentally on the atomic scale by means of $\mu^+$SR.%, even if the diffusion coefficient is not directly measurable in the different directions. 

Anisotropy of Li diffusion in LiFePO$_4$ was also reported by AC impedance spectroscopy measurements on single crystal \cite{Jiying2008}%, which was made identically to the pieces used here. 
The anisotropy was inferred from the difference in the observed spectra along each crystalline a-xis and the reduced activation energy along the b-axis with $E_a= 540$~meV, compared to the other two. Such value is significantly higher than what is obtained at this work and from powder $\mu^+$SR \cite{Sugiyama2011}. The difference in the absolute value can be attributed to each experimental technique sensitivity. While AC impedance spectroscopy suggest static behaviour even at 400~K, a clear dynamical internal field distribution is observed in this and previous powder measurement \cite{Sugiyama2011}. A difference in the absolute value is common to see while comparing values obtained from macroscopic techniques and microscopic techniques like $\mu^+$SR. Regardless though, anisotropic Li diffusion is confirmed. 

Recent $\mu^+$SR measurement on LiFePO$_4$ nanocrystalline powders suggested that bulk diffusion is insignificant \cite{Benedek2020} and the observed Li ion diffusion is rather originating from the surface region. It was proposed that Li-ion diffusion observed in previous powder LiFePO$_4$ studies was originating from surface effects \cite{Benedek_2019}, attributed to small powder sizes. Contrary, our follow-up experiment provides a clear proof of bulk diffusion. In fact, the obtained diffusion values here are similar to the results obtained from $\mu^+$SR powder measurements on LiFePO$_4$ \cite{Sugiyama2011}. Since we may safely neglect surface effects in our single crystal measurements, we may conclude that bulk diffusion is present in LiFePO$_4$. The discrepancy between these two results may find its explanation in the difference in sample synthesis protocols and/or the size of each individual grain/crystal. Evidently, the synthesis routes for nano-crystals \cite{Benedek_2019,Benedek2020} and large single crystals \cite{Li2006} are very different. It is therefore very likely that the resulting samples also could have differences in, e.g. local defects, impurities, or local disorder and strains, which all could clearly influence ion diffusion. Further comparative investigations of single crystal and nano-crystalline samples using diffraction and/or total scattering methods will be valuable to further clarify this important question.

%suggesting that bulk diffusion is also present in powder form. Therefore, we may conclude that the diffusion mechanism in LiFePO$_4$ is indeed interstitial mechanism, supported by the the obtained diffusion coefficient in this study, assuming interstitial mechanism, and its agreement with the theoretical prediction \cite{Morgan2004}. To clarify, our study show bulk diffusion is the significant contribution, but the surface may still have a small role

Finally, we wish to discuss the future possibilities of utilising these kind of measurements in other type of compounds. Anisotropic ion diffusion can clearly be detected using the $\mu^+$SR technique. As shown, Li nuclear moment fluctuations perpendicular to the initial muon spin polarisation results in higher fluctuation rate in the data. Naturally, any ion posing nuclear moments, $e.g.$ Na, are prime candidates. In fact, any compound that presumably exhibits anisotropic ion diffusion would benefit from measurements presented here. One prime example includes NaCoO$_2$, which is known to expand the diffusion path from 1D (above 295 K) to 2D (above 400~K) \cite{Medarde2013}. 

%For the sake of completeness, we will discuss the validity of utilising a KT function for fitting single crystal data \cite{Solt1995}. Depending on the muon site, the value of $\Delta$ along each crystalline axis may be different enough that the KT function would not valid, since it assumes an isotropic distribution along each perpendicular axis. Naturally, the value of $\Delta$ along each axis depends on the site in question. We have thus calculated the expected powder $\Delta$ and the expected values along each direction for all muon sites proposed by Ref.~\cite{Sugiyama2011}. Apart from site (0.0416,0.2500,0.9172), the the expected $\Delta$ along each crystalline axis are quite similar. The small differences are probably small enough such that the KT approximation is still valid. This is supported by the quality of our fits (Fig.~\ref{fig:_ZFLFSpec}) and by the fact that the obtained $\Delta$ along each direction resulted in similar values (Fig.~\ref{fig:_Para}(b))

\section{\label{sec:conclusion}Conclusions}

$\mu^+$SR is a relatively new method to measure diffusion of ions at the atomic scale. It is especially valuable for materials containing transition metals, which are challenging for NMR. In this work we deepened the understanding of this method by exploiting anisotropic diffusion in a single crystal, namely in LiFePO$_4$. Fortunately, the calculated internal magnetic field distribution at the muon sites was similar in all directions, and therefore we could fit the data with the static / dynamic GKT function, like it is done for powders. Similarly to powders, we observed bulk Li diffusion above $\sim$150K. We found clear differences in the direction dependent hopping rates, $\nu$, which we ascribed not to the different jump probability of the Li ion in the different directions, but to the change in transition probability between the two Zeeman states of the muon. Measurements on the small crystals limited the data quality, but
the obtained diffusion coefficient for this compound, $D^{\rm protocol~2}_{Li}(300~$K$)=1.23\times10^{-9}$~cm$^{2}$/s, is closest to what has been estimate with first principle calculations to this date. This is naturally attributed to the fact that powder average will inevitably lower the measured fluctuation rates and thus the estimated diffusion coefficient will be 'artificially' detected as smaller in materials with anisotropic diffusion. 

%One should however note that utilisation of KT for the data fit cannot be given for granted. For the title compound however, the second moment in each direction for the present compound was shown to be quite similar.

\begin{acknowledgments}
This research was supported by the Swedish Research Council (VR) via a Neutron Project Grant (Dnr. 2016-06955), the Carl Tryggers Foundation for Scientific Research (CTS-18:272) and the Swedish Foundation for Strategic Research (SSF) within the Swedish national graduate school in neutron scattering (SwedNess). Y.S. is funded by VR through a Starting Grant (Dnr. 2017-05078) as well as the Chalmers Area of Advance-Materials Science. J.S. acknowledges support from Japan Society for the Promotion Science (JSPS) KAKENHI Grants No. JP18H01863 and No. JP20K21149. All images involving crystal structure were made with the VESTA software \cite{Momma2008} and the $\mu^+$SR data was fitted using \texttt{musrfit} \cite{musrfit}.

%This research was supported by Marie Sk{\l}odowska Curie Action, International Career Grant through the European Union and Swedish Research Council (VR), Grant No. INCA-2014-6426 as well as the Ministry of Education, Culture, Sports, Science and Technology (MEXT) of Japan, KAKENHI Grant No. 23108003 and Japan Society for the Promotion Science (JSPS) KAKENHI Grant No. 26286084. Finally, XX also acknowledge research funds received from the Wenner-Gren Foundations. 
\end{acknowledgments}

%------------------------------------ REFRERENCES
\bibliography{Refs} % Call your Name.bib file with all the references.
\end{document}